\documentclass{article}
\usepackage{spconf,amsmath,graphicx}
\usepackage{hyperref, booktabs, multirow}
\usepackage[subtle]{savetrees}

\title{Content Leakage in LibriSpeech and Its Impact on the Privacy Evaluation of Speaker Anonymization}
\name{
Carlos Franzreb\textsuperscript{1}, Arnab Das\textsuperscript{1}, Tim Polzehl\textsuperscript{1}, Sebastian Möller\textsuperscript{2}
\thanks{Funded by Federal Ministry of Education and Research, Germany (BMBF 16KIS2048).}
}
\address{
\textsuperscript{1} DFKI, Germany,
\textsuperscript{2} Technical University of Berlin, Germany
\\ \url{carlos.franzreb@dfki.de}
}
\begin{document}
%\ninept
%
\maketitle
\begin{abstract} % 100 to 150 words
Speaker anonymization aims to conceal a speaker's identity, without considering the linguistic content.
In this study, we reveal a weakness of Librispeech, the dataset that is commonly used to evaluate anonymizers: the books read by the Librispeech speakers are so distinct, that speakers can be identified by their vocabularies.
Even perfect anonymizers cannot prevent this identity leakage.
The EdAcc dataset is better in this regard: only a few speakers can be identified through their vocabularies, encouraging the attacker to look elsewhere for the identities of the anonymized speakers.
EdAcc also comprises spontaneous speech and more diverse speakers, complementing Librispeech and giving more insights into how anonymizers work.
\end{abstract}
\begin{keywords}
Speaker anonymization, privacy, speaker recognition
\end{keywords}
\section{Introduction}
\label{sec:intro}

% spkanon
Given an utterance, speaker anonymization aims to conceal the source speaker's identity while keeping the linguistic and paralinguistic content intact \cite{jin_speaker_2009}.
% Privacy evaluation
The commonly used privacy evaluation was designed by the VoicePrivacy initiative \cite{tomashenko_introducing_2020}.
It simulates an attack on the privacy of anonymized speakers, where the attacker is assumed to have access to the anonymizer.

% TODO: there is a x-tick hanging in the left figure, and the same numbers are already plotted in the phone duration results - it could be removed.
\begin{figure*}[t]
\minipage{0.32\textwidth}
    \includegraphics[width=\linewidth]{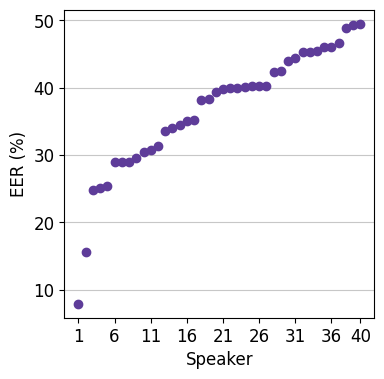}
    \caption{Speaker EERs for Librispeech anonymized with STT-TTS.}
    \label{fig:stt_tts_speaker_eers}
\endminipage
\hfill
\minipage{0.32\textwidth}
    \includegraphics[width=\linewidth]{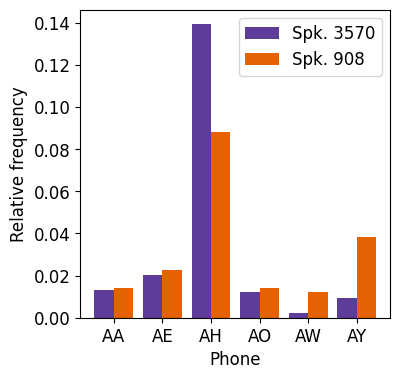}
    \caption{Relative phone frequencies for two Librispeech speakers.}
    \label{fig:phone_counts}
\endminipage
\hfill
\minipage{0.30\textwidth}
    \includegraphics[width=\linewidth]{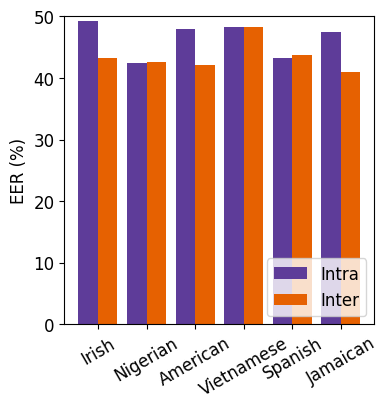}    
\caption{Intra- and inter-EERs for EdAcc accents, anonymized with STT-TTS.}
\label{fig:edacc_stt_tts_demographics}
\endminipage
\hfill
\end{figure*}

% Librispeech
Librispeech \cite{panayotov_librispeech_2015} is used both for training (\textit{train-clean-360}) and evaluation (\textit{test-clean}).
It is derived from audiobooks in the LibriVox project, with different speakers usually reading from different books.
% Advantages
Librispeech is one of the most popular datasets in speech processing, bridging the gap between numerous fields.
Its lack of noise and speaker diversity allows us to assess model performance under optimal conditions, estimating the potential of the models, but ignoring their robustness.
% Drawbacks
Librispeech has other drawbacks for speaker anonymization, like that it is split at silence intervals, hindering the attacker from leveraging long-term speech characteristics such as prosody \cite{zen_libritts_2019}.

% Contributions
In this study, we discover a new weakness of Librispeech with respect to the privacy evaluation: the content of the books read by each speaker in the Librispeech dataset is so distinct, that the speakers can be identified by only leveraging the content.
This is problematic for speaker anonymization, where linguistic information is ignored: what the speakers say is not anonymized, and it should not be leveraged by the evaluation's attacker to identify speakers.

% EdAcc
EdAcc \cite{sanabria_edinburgh_2023}, which has also been proposed for the privacy evaluation \cite{franzreb_comprehensive_2023}, comprises dyadic conversations between 60 pairs of speakers with different backgrounds and accents.
Its advantages over Librispeech is that it's spontaneous speech, which differs from read speech in terms of prosody and emotion, and features L2 speakers.
% Contribution
We show that EdAcc does not leak speaker identity through the content as much as Librispeech, making it a great addition to the privacy evaluation, and also highlight the value of its population segments.

\section{Privacy evaluation}

% evaluation framework
We run our experiments with the SpAnE framework \cite{franzreb_comprehensive_2023}, which implements the attacker from the VPC 2024 privacy evaluation \cite{tomashenko_voiceprivacy_2024}.
The only difference is the size of the speaker recognizer: SpAnE uses the standard ECAPA-TDNN provided by SpeechBrain \cite{dawalatabad_ecapa-tdnn_2021}, whereas the VPC 2024 halves the number of channels.
% Evaluation data
The split between trial and enrollment utterances for the evaluation is the one proposed in \cite{franzreb_optimizing_2025}, with 20 utterances per speaker in each split.

% Evaluation
The attacker's task is to identify the speakers of the trial utterances, which are anonymized, by comparing them with the enrollment utterances.
The attacker trains the recognizer with anonymized speech, and also anonymizes the enrollment utterances to reduce the domain mismatch, as their speaker embeddings are extracted with the same recognizer.
The enrollment speaker embeddings are averaged across speakers, and then compared with the embeddings of the trial utterances with cosine distance.
% EER
The equal error rate (EER) \cite{maouche_comparative_2020} is used to assess the effectiveness of the attack.
$EER=0\%$ means that the speakers of all trial utterances could be identified, whereas $EER=50\%$ implies perfect anonymization, as the attacker is making binary decisions randomly.

\subsection{Anonymizer}

% Anonymizer
As the anonymizer, we use a speech-to-text-to-speech (STT-TTS) pipeline.
It transcribes the source speech with the speech recognizer Whisper-small \cite{radford_robust_2022} and synthesizes the transcript with the multi-speaker TTS pipeline from NeMo\footnote{\url{https://catalog.ngc.nvidia.com/orgs/nvidia/teams/nemo/models/tts_en_multispeaker_fastpitchhifigan}}, which comprises a FastPitch model \cite{lancucki_fastpitch_2021} to generate a spectrogram from text and a HiFiGAN synthesizer \cite{kong_hifi-gan_2020}.
% Why
STT-TTS is a perfect anonymizer: the transcription step removes all speaker information besides the information present on the linguistic channel, which is ignored by our working definition of privacy.
The timbre, the pitch contour and the phone durations of the input speech are not present in the transcript, and are therefore not present in the synthesized speech either.
% What this means for the privacy evaluation
Therefore, if speakers anonymized with STT-TTS are identified by the attacker, it must be because of what they are saying.

\section{Related work}

Several studies have investigated the role of phone durations in the privacy evaluation \cite{jin_speaker_2009, tomashenko_analysis_2025, tomashenko_exploiting_2025, franzreb_private_2025}.
\cite{tomashenko_analysis_2025} investigates how identity is encoded in phone durations, and whether it can be exploited to identify anonymized speakers.
They design two metrics for comparing the speakers of different utterances based on the mean durations of each phone.
When 60 utterances are used to compute the mean phone durations, these metrics can identify some speakers that are anonymized with an STT-TTS anonymizer ($EER=26\%$), although the original phone durations are not preserved.

The results were surprising, as the 
anonymizer alters phone durations and is expected to provide perfect privacy.
When evaluated with the VPC 2024 \cite{tomashenko_voiceprivacy_2024}, its privacy is indeed estimated to be perfect ($EER=48.2\%$).
The authors hypothesize that this occurs because of how the book content affects the speaking style.
A recent study \cite{gaznepoglu_you_2025} reaches a similar conclusion after identifying Librispeech speakers through their content with BERT \cite{devlin_bert_2019}.
In this study, we provide further evidence supporting this hypothesis.
We also link this finding with the conventional privacy evaluation: by increasing the size of the recognizer, as done by SpAnE \cite{franzreb_comprehensive_2023}, the EER decreases to $34.8\%$.
The larger recognizer is able to identify speakers by their vocabularies, if that is the only way.

\section{Privacy evaluation with Librispeech}

For the privacy evaluation, we can measure how easy it is to identify each speaker with its speaker-specific EER.
The EER for speaker $A$ is computed by only considering the pairs of trial and enrollment utterances where $A$ is the speaker of the trial utterance.
Figure \ref{fig:stt_tts_speaker_eers} depicts the speaker EERs for the Librispeech speakers anonymized with STT-TTS.
Some speakers have much lower EERs than others, meaning that they are easy to discriminate from the rest.
If our hypothesis is true, there should be a correlation between how distinctive the content of each speaker's book is, and the speaker EERs.

\subsection{Phone counts in transcripts}

% Experiment
This first experiment investigates whether the phone frequencies in the Librispeech books are useful for discriminating speakers.
We first transform the evaluation dataset's transcripts into phone sequences with an existing grapheme-to-phoneme (g2p) model \footnote{\url{https://github.com/Kyubyong/g2p}}.
The phonetic alphabet comes from the CMU pronouncing dictionary\footnote{\url{http://www.speech.cs.cmu.edu/cgi-bin/cmudict}}, without considering stress; it comprises 39 phones.
We then count the phones for each speaker, and divide them with the total number of phones per speaker.
Figure \ref{fig:phone_counts} shows the frequencies of two speakers for the phones that start with ``A''.
This already shows some differences between the two speakers: speaker 3570, who has the lowest speaker EER ($7.9\%$), has a higher frequency for the phone ``AH'', whereas speaker 908 ($EER=33.5\%$) has a higher frequency for phone ``AY''.

% Cosine distances
We measure cosine distances between the relative phone frequencies of each pair of speakers.
The average cosine distance of each speaker across pairs is a measure of its distinctiveness: a higher distance means that the speaker's relative phone frequencies are more different from those of the other speakers.
% Results
These distances correlate with the speaker's EERs with a Pearson correlation coefficient of $r=0.59$.
Speaker 3570, whose EER was the lowest, also has the highest avg. distance.
This speaker is reading from the book \textit{The theory of the leisure class} \cite{veblen_theory_1899}, published in 1899.

% Conclusion and segway
This simple experiment, which does not require training nor takes into account the order and duration of the phones, shows that what the speakers are reading already provides the attacker with clues to discriminate them.
The large recognizer from the privacy evaluation is able to model much more complex features than the ones used here.

\subsection{Phone durations}

% Tomashenko's Interspeech paper
To perform a stronger privacy attack, we replace the mel-spectrogram which is usually fed to the speaker recognizer with a representation that only encodes phones and their durations.
This representation is a one-hot encoding \cite{hancock2020survey} of the phonetic transcript.
Instead of a 1, the numbers in the one-hot vectors are the durations of their respective phones.
This representation is formalized in Equation \ref{eq:phone_durations}: $R_{i,t}$ is the resulting value in the representation matrix $R$, which is set to the phone's duration $d_t$ only when the vocabulary index $i$ matches the current phone's index $\phi(p_t)$, a condition enforced by the Kronecker delta ($\delta$).
In this way, the recognizer can only leverage the sequence of phones and their durations to identify speakers.
All the other aspects of speech which contribute to speaker identity, like timbre and pitch, are not present in this representation.

\begin{equation}
    R_{i,t} = d_t \cdot \delta_{i, \phi(p_t)}
    \label{eq:phone_durations}
\end{equation}

% Tomashenko vs. ours
\cite{tomashenko_exploiting_2025} proposed and used this approach to show that phone durations can be used to identify Librispeech speakers when the durations are preserved.
Here, we show that this is still the case when the durations are not preserved, because the underlying content is distinctive enough to discriminate speakers regardless.

% Our implementation
We use the phone recognizer from private kNN-VC \cite{franzreb_private_2025}, which was trained on Librispeech \textit{train-clean-100} and achieves an accuracy of 92\% and a phone error rate (PER) of 2\% on a held-out test set.
The phone recognizer is used to predict a sequence of phones from input speech, including their durations.
It uses the same phonetic alphabet as before, with the addition of silence.
The rest of the privacy evaluation is exactly the same as before.

% Experiments
We test this approach both with original speech and the STT-TTS anonymizer.
The results are presented in Table \ref{tab:phone_durations_results}, compared with the results of the conventional privacy evaluation, where mel-spectrograms are used as features instead of the phone durations.
When using mel-spectrograms, speakers that have not been anonymized can be identified very accurately ($EER=0.4\%$).
Using phone durations instead of mel-spectrograms degrades the efficacy of the attack, but it is still able to identify many speakers ($EER=23.7\%$).
The increase is expected, as many other speech characteristics can be leveraged to identify speakers (e.g. timbre, pitch).
% STT-TTS
When the utterances are anonymized with the STT-TTS anonymizer, using phone durations as features leads to a comparable EER than with the mel-spectrograms ($34.8\%$ vs $34.5\%$).
The similarity of the outcomes suggest that all the speaker identity that is captured by the conventional attacker comes from the phone durations.
Speaker 3570 is again the easiest to identify for the recognizer trained with phone durations ($EER=10.1\%$).

\subsection{Phones}

Even when the durations are not preserved, several speakers can still be identified.
It is therefore possible that the phones themselves are leaking their identity, and not their durations.
To test this, we run the same experiment as before, but without the durations: the duration of each phone in the representation is set to 1.
The results for this experiment are depicted in Table \ref{tab:phone_durations_results}.
For the original Librispeech dataset, removing the phone durations increases the EER from $23.7\%$ to $30.4\%$; when anonymized with STT-TTS, the EER decreases from $34.5\%$ to $32.3\%$.
The results for the original and anonymized utterances are similar because, except for the transcription errors made by the speech recognizer, the two utterances share the same phone sequence.
For the anonymized utterances, ignoring the phone durations improves evaluation performance, implying that, when anonymized, phone durations are not helpful for identifying speakers.
The content does not have a sufficient influence on the durations enough for it to remain after the transcription.

% Conclusion
The experiments in this section show that the content of the books read by the Librispeech speakers are too distinctive.
This fact interferes with the privacy evaluation, and should be considered when interpreting privacy estimates.
% STT-TTS
STT-TTS is a perfect anonymizer for our task; it should therefore be considered as the upper bound for the achievable privacy.
% Segway
In the next section, we will evaluate the privacy of EdAcc, which is less distinct in terms of content.

\begin{table}[]
    \centering
    \caption{EERs as percentages for the two evaluation datasets, depending on the feature extractor. The asterisk in the last rows means that the recognizer is trained with Librispeech and EdAcc utterances.}
    \label{tab:phone_durations_results}
    \smallskip
    \begin{tabular}{cccc}
        \toprule
        \textbf{Dataset} & \textbf{Features} & \textbf{Original} & \textbf{STT-TTS} \\
        \midrule
        \multirow{3}{*}{Librispeech} & Mel-spectrograms & 0.4 & 34.8 \\
         & Phones + durations & 23.7 & 34.5 \\
         & Phones & 30.4 & 32.3 \\
        \midrule
        \multirow{3}{*}{EdAcc} & Mel-spectrograms & 6.5 & 45.9 \\
         & Phones + durations & 39.0 & 45.0 \\
         & Phones & 42.1 & 48.5 \\
        \midrule
        EdAcc* & Mel-spectrograms & 3.0 & 43.3 \\
        Librispeech* & Mel-spectrograms & 0.4 & 35.6 \\
        \bottomrule
    \end{tabular}
\end{table}

\section{Privacy evaluation with EdAcc}

% EdAcc pros, but small size
As EdAcc \cite{sanabria_edinburgh_2023} comprises spontaneous conversations, the content of each speaker's utterances should not be as uniform and distinctive as in Librispeech.
% Small size
EdAcc is much smaller than Librispeech, with 22 hours of speech in total, evenly divided into two sets (\textit{dev} and \textit{test}).
EdAcc \textit{dev} comprises 62 speakers, and EdAcc \textit{test} 60 speakers.
% SpAnE
For training the recognizer, we follow the approach of \cite{franzreb_comprehensive_2023}, which trains it on Librispeech \textit{train-clean-360} and EdAcc \textit{dev}, to overcome the small size of EdAcc.
We compare this approach with recognizers trained only on Librispeech.

% Evaluation data
EdAcc \textit{test} is used as the evaluation data, without considering the utterances shorter than 2 seconds or longer than 30 seconds.
The dataset is heavily imbalanced in terms of utterances per speaker: on average, there are 78 utterances per speaker, but the standard deviation is 73.
Experiments with Librispeech \cite{franzreb_optimizing_2025} suggest that 20 trial and 20 enrollment utterances per speaker are enough to ensure reliable results; having more did not improve the effectiveness of the privacy attack.
To account for differences between the datasets, we keep 60 utterances per speaker; 20 of them are used for the enrollment set, and the rest for the trial set.
% Speakers with less than 30 utterances
The utterances of the 9 speakers with less than 30 utterances are split evenly between the trial and enrollment sets.

% Results
Table \ref{tab:phone_durations_results} shows the EERs for the EdAcc evaluation dataset, both for original speech and the STT-TTS anonymizer.
For the middle rows, the recognizers are trained only with Librispeech.
The last two rows depict the EERs when the recognizers are trained with both datasets.
Doing so leads to a stronger attack for the EdAcc evaluation data.
This recognizer can still identify Librispeech speakers effectively, meaning that one recognizer can be used to evaluate both datasets, simplifying the evaluation.

% Original speech
When mel-spectrograms are used as features, the EER for the original speech increases from $0.4\%$ for Librispeech to $6.5\%$ for EdAcc.
Adding EdAcc \textit{dev} to the recognizer's training data reduces EdAcc \textit{test}'s EER to $3.0\%$.
% STT-TTS
When the utterances are anonymized with STT-TTS, EdAcc's EER is $45.9\%$ when the recognizer is trained on Librispeech, and $43.3\%$ for the recognizer trained on both datasets.
As the recognizer can only leverage phones and their durations to identify the anonymized speakers, these results suggest that the content of the EdAcc utterances does not leak speaker identity as much as Librispeech.

% Comparison with phone durations
To assess EdAcc's content leakage, we run the same experiment from the last section, with the recognizers trained on Librispeech.
The EERs for the original data and STT-TTS can be found in Table \ref{tab:phone_durations_results}.
% original
As for Librispeech, the EdAcc EER of the original speech is much larger for phone durations ($39.0\%$) than for mel-spectrograms ($6.5$).
This EER is also larger than its Librispeech analog ($23.7\%$).
% STT-TT
For the STT-TTS anonymizer, EdAcc's EER is comparable for the two features: $45.9\%$ when mel-spectrograms are used, and $45.0\%$ when phone durations are used.
The recognizer is capable of identifying a few speakers based on the phonetic characteristics of what they are saying, but to a lesser degree than for Librispeech ($EER=34.5\%$).

\subsection{Privacy of population segments}

% How we did it in the intelligibility evaluation
Another advantage of EdAcc is its rich metadata, which tells us how much privacy each anonymizer provides for different population segments.
It includes the accent of the speakers, when they started learning English, their ethnicity, their age and many other relevant facts.
% How we do it for privacy
Inspired by \cite{tayebi_arasteh_addressing_2024}, we propose to measure two aspects for each population segment:

\begin{enumerate}
    \itemsep -0.3em
    \item \textbf{Intra-EER}: what is the privacy provided to the speakers of a certain population segment between themselves?
    \item \textbf{Inter-EER}: what is the privacy provided to the speakers of a certain population segment with respect to all speakers?
\end{enumerate}

% Why are both important
The intra-EER looks at how distinguishable individuals are within the same demographic group.
Assuming that the attacker knows this demographic attribute about the speaker that it is trying to identify, can we protect their identity?
In contrast, the inter-EER measures the risk of being identified for speakers of each population segment.
A fair anonymizer should protect the identity of all segments equally.
Also, if the privacy estimate for a population segment is considerably worse than for the rest, it is likely that the attacker can infer the population segment, which constitutes sensitive information.
% In practice
In practice, the intra-EER considers only the speakers of each demographic group, whereas these speakers are the trial speakers for the inter-EER, whereas the whole dataset's speakers are the enrollment speakers.

% Which results we look at here
To illustrate how it works, we consider the speaker's accents and the STT-TTS anonymizer.
STT-TTS defines how much privacy can be achieved for each accent without altering what is being said.
We only consider accents that comprise at least three speakers.
% Results
Figure \ref{fig:edacc_stt_tts_demographics} shows the results for the accent of the speakers.
Speakers with either a Spanish or a Nigerian accent are relatively easy to identify, both with respect to other groups and between themselves, with intra- and inter-EERs around $43\%$.
Speakers with an Irish, American or Jamaican accent are easy to discriminate from speakers with other accents, with inter-EERs around $42\%$, but harder to discriminate among themselves, with intra-EERs above $47\%$.
These results suggest that speakers with these three accents have distinct vocabularies, which allow the recognizer to discriminate them from speakers with other accents, but it does not help to discriminate them among themselves, as they share the same vocabulary.
Such an outcome is expected from different communities of native speakers \cite{holmes_introduction_2013}.

\section{Conclusion}

Librispeech is the most popular dataset for evaluating anonymizers.
The lack of speaker diversity and the fact that the speech is read, not spontaneous, make this dataset a poor choice to estimate an anonymizer's performance for real use cases.
Our study shows that the attacker can identify several speakers just by focusing on the content, as the books vary greatly in their vocabulary.
We propose to include EdAcc, whose speech is spontaneous and comes from diverse speakers.
Its content cannot be leveraged to identify speakers, which encourages the recognizer to focus on the aspects of speech that we are interested in (e.g. timbre and prosody).
% For anonymizers
For imperfect anonymizers, EdAcc may characterize weaknesses for particular population segments, which is not possible with Librispeech.

\vfill\pagebreak

% References should be produced using the bibtex program from suitable
% BiBTeX files (here: strings, refs, manuals). The IEEEbib.bst bibliography
% style file from IEEE produces unsorted bibliography list.
% -------------------------------------------------------------------------
\bibliographystyle{IEEEbib}
\bibliography{strings,refs}

\end{document}